\documentstyle[preprint,aps]{revtex} 
\begin{document} 
\draft 
\title{Coherent Population Trapping with Losses on the Sodium D$_1$ Line Hanle
Effect}

\author{F. Renzoni, W. Maichen,  L. Windholz and E. Arimondo $^{(1)}$}

\address{\ Institut f\"ur Experimentalphysik, Technische Universit\"at
Graz, A-8010 Graz, Austria}  

\address{\ $^{(1)}$ Unit\`a Istituto Nazionale di Fisica della Materia, Dipartimento di
Fisica dell'Universit\`a, I-56126 Pisa, Italia}

\date{\today{}}

\maketitle

\begin{abstract}
We consider the coherent population trapping phenomenon in a
thermal sodium atomic beam.
We compare the different coherent population trapping schemes that can be
established on the D$_{1}$ line using the Zeeman sublevels of a
given ground hyperfine state. The coherent population
trapping preparation is examined by means of a Hanle effect configuration.
 The efficiency of the coherent population trapping phenomenon has been examined
in presence of optical pumping  into hyperfine levels
external to those of the excited transition. We show that both the contrast and  the
width of the coherent population trapping resonance  strongly decrease when 
the optical pumping  rate is increased. In the experiment, the loss
rate due to optical pumping has been controlled by means of a laser
repump of variable intensity.\\
\end{abstract}

\section{Introduction}

Recently the Coherent Population Trapping (CPT) phenomenon
has received a large interest, also in connection with its applications,
for example the laser cooling below the recoil limit or lasing
without inversion (see \cite{ari1} for a review). At the beginning 
CPT theoretical studies were restricted to the three-level $\Lambda$-system 
\cite{ari2,gray}, but recently  they  have been
extended to different atom-light interaction schemes \cite{smir,papoff}.
Despite the large amount of data available on the CPT phenomenon
exploited in several atom-light interaction schemes, the features related to
different level schemes and established on  a given atomic species, have never been
directly compared. On the other hand, the increasing interest for CPT 
applications requires realistic calculations  taking in account
processes as the loss towards the external states, the Doppler broadening of the
absorbing transition, the collisions, all of them present in the experiment and
determining the strength of the CPT resonance.

The main goal of the present investigation is to study the
dependence of the CPT phenomenon on the atom-laser interaction
parameters. In this paper we focus our attention on the  level schemes
that can be established on the sodium $D_1$ line using as ground states the Zeeman
sublevels of the same hyperfine component. We investigate theoretically and
experimentally the CPT features associated to the different hyperfine
optical transitions.
As shown in Fig. 1, the hyperfine transitions composing the D$_{1}$ line are:
$F_{g}=1\rightarrow F_{e}=1$, $F_{g}=2\rightarrow F_{e}=1$,
$F_{g}=2\rightarrow F_{e}=2$ and $F_{g}=1\rightarrow F_{e}=2$.
Within all those optical transitions, only a limited number
of Zeeman sublevels contribute to the preparation of the CPT coherent
superposition of states, and precisely those
connected by heavy lines in Fig. 1. For instance the $F_{g}=1\rightarrow
F_{e}=1$ transition contains a $\Lambda$ system that excited by
$\sigma^{+},\sigma^{-}$ light produces the coherent superposition of
ground states not interacting with the laser radiation. The
$F_{g}=2\rightarrow F_{e}=2$ transition contains a M system that again excited
by $\sigma^{+},\sigma^{-}$ light produces a coherent superposition of 
three ground states not interacting with the laser radiation. For the
transition $F_{g}=2\rightarrow F_{e}=1$ both the coherent dark
superpositions listed above for the $F_{g}=1\rightarrow F_{e}=1$ and 
$F_{g}=2\rightarrow F_{e}=2$ transitions are present. 
Finally the transition $F_{g}=1 \rightarrow F_{e}=2$ is not relevant 
for CPT in the $\sigma^{+},\sigma^{-}$ light configuration, because it does not contain
a coherent superposition noncoupled to the laser field.

The experiment is based on the laser excitation of a sodium atomic beam. To
produce CPT in the sodium atoms and to investigate its production we have used a
Hanle effect configuration. The sodium atoms are excited by monochromatic
linearly polarised laser light resonant with an hyperfine optical transition; the
degeneracy of the ground state Zeeman sublevels is removed by the introduction of
an external magnetic field parallel to the laser propagation direction. For zero
magnetic field the atomic system is optically pumped into a coherent
superposition of ground states non interacting with the laser field, i.e., the  dark
or noncoupled state. For a fixed laser frequency, scanning the external magnetic
field around the zero value, the atomic fluorescence emitted at right angles with
respect to the directions of the laser field propagation and of the magnetic field
exhibits a minimum at zero magnetic field with a lineshape dip typical of the Hanle/CPT
phenomenon.

 We have measured the contrast and the linewidth of the resonance in the
Hanle/CPT lineshape versus the intensity of applied laser field. These measurements are
compared to analytical and numerical analyses.  For a closed optical transition 
the CPT process is quite straightforwardly  understood and 
described in the frame of the optical Bloch equations. With an atomic sample
in an initial uniform Zeeman distribution, the atomic preparation into the
coherent superposition increases with a time constant determined by the optical
pumping rate into the noncoupled state.  For an  open optical transition, i.e., in the
presence of atomic levels external to the hyperfine transition resonant with the
laser light, the atomic preparation into the coherent superposition is modified by
the optical pumping into those levels. The atomic time evolution is governed by the
competition between optical pumping into the noncoupled state and the optical
pumping into the external  hyperfine states.

 We have
investigated, both theoretically and experimentally, how the contrast and linewidth
features of the CPT resonance depend on the optical pumping rate towards the
external levels. In effect for the different hyperfine optical transitions of the
sodium D$_1$ line, owing to their different optical pumping rates, the  CPT
resonances have different linewidths and contrasts. The optical pumping rate
into external levels may be compensated by the application of a repumping
laser of variable intensity. Thus we have studied the CPT features 
of a given scheme of hyperfine levels as a function of the external
optical pumping loss rate by applying a repumping laser of variable intensity.
Some of our results, as the narrow linewidth realized on open
transitions, can be applied to the magnetometry based on coherent population
trapping recently introduced by Wynands {\it et al} \cite{wynands}.

Our investigation should be compared to previous CPT studies. The more recent
and detailed investigation of the CPT resonance has been performed by Ling {\it et al}
\cite{xiao}. Previous accurate studies of the Hanle effect in the ground state
\cite{cohen,rasmussen,walther,brand,picque}  have examined the lineshape
dependence on the laser intensity.   However none of those  studies
analysed the role of the different hyperfine transitions as presented
here.  Ling {\it et al} \cite{xiao} have examined the role of the
Doppler broadening on  CPT. Their results show that in our experiment
the  Doppler broadening associated to the residual divergence of the
sodium atomic beam has a negligible influence on the measured contrast
and linewidth. Thus we have not included the Doppler broadening in our
analysis. In the comparison between our data and the theoretical analysis
we have discovered that our measured contrast could not reach the
theoretical value because the magnetic field compensation was not
accurate as required. On the contrary in the experiment by
Picqu\'e\cite{picque}  the very good magnetic field compensation allowed
the author to reach the contrast predicted by the theory. Thus we have
used the data of ref. \cite{picque} for completing the comparison with
our theoretical analysis.
	
In the present work Section II contains a theoretical analysis of the
CPT process based on the analytical and numerical solution of the optical Bloch
equations. Section III describes the experimental setup and the experimental results.
Section IV contains the comparison between the theoretical analysis and the
experimental results. In Section V conclusions are presented.

 \section{Optical Bloch Equation}

We consider a sodium atom interacting with a linearly polarized
monochromatic laser light resonant with one of the hyperfine transition
of the $D_{1}$ line and propagating in the direction $Oz$
\begin{equation}
\vec{E}(z,t) = {{\cal E}\over 2}\vec{\epsilon}_x e^{i(k z - \omega t)}+
c.c. = {\sqrt{2}{\cal E}\over 4}\left( \vec{\epsilon}_{\sigma^+}+ \vec{\epsilon}_{\sigma^-}
\right) e^{i(k z - \omega t)}+ c.c.
\end{equation}
with $\vec{\epsilon}_i$ the unit vector of the $i$ polarization. We indicate by
$F_g\rightarrow F_e$ the transition pumped by the laser, $F_g$ and $F_e$ being the
quantum numbers of the total angular momentum of the hyperfine levels of the $^2
S_{1/2}$ and  $^2 P_{1/2}$ levels, respectively. The quantum number of the total
angular momentum of the other hyperfine level of the  $^2 S_{1/2}$ level will be
denoted by $F_{g^{'}}$. A magnetic field $B$ is applied in the direction $Oz$.

For the $z$-axis as quantization axis, the optical Bloch
equations (OBE) for the system $F_g \rightarrow F_e$ under examination 
have the following form ($|e_j> = |JIF_e j>, |g_j> = |JIF_g j>$):
\begin{mathletters}
\begin{eqnarray}
\dot{\rho}_{e_i e_j} &=& - [ i \omega_{e_i e_j} + 
\Gamma_{F_e \to F_g;} (1+\alpha_{F_e \to F_g; F_{g^{'}}})]
\rho_{e_i e_j}+{i\over \hbar}\sum_{g_k}\left( \rho_{e_i g_k}  V_{g_k
e_j}-
V _{e_i g_k}\rho_{g_k e_j} \right) \label{bloch1} \\
\dot{\rho}_{e_i g_j} &=& - \left[ i\omega_{e_i g_j}+
\frac{\Gamma_{F_e \to F_g}(1+\alpha_{F_e \to F_g; F_{g^{'}}})}{2} \right]
\rho_{e_i g_j}+{i\over \hbar}\left( \sum_{e_k}\rho_{e_i e_k}  V_{e_k
g_j}-
\sum_{g_k}  V _{e_i g_k}\rho_{g_k g_j} \right) \label{bloch2} \\
\dot{\rho}_{g_i g_j} &=& - i\omega_{g_i g_j} \rho_{g_i g_j} +
{i\over \hbar}\sum_{e_k} \left(  \rho_{g_i e_k} V_{e_k g_j}-V _{g_i e_k}
\rho_{e_k g_j} \right) + \left( {d\over dt}\rho_{g_i g_j}
\right)_{SE}  
\label{bloch3}
\end{eqnarray}
\end{mathletters}

 The quantities $\omega_{\alpha_i,\beta_j}$, with
$\alpha , \beta = (e,g)$,
 represent the frequency separation between the levels $\alpha_i$ and $\beta_j$,
including the Zeemand splittings of the ground and excited levels due to the applied
magnetic field B 
\begin{equation}
\omega_{\alpha_i,\beta_j} = {E_{\alpha_i}-E_{\beta_j}\over \hbar}.
\end{equation}
$\Gamma$ is the total spontaneous emission rate for any excited
level, $\Gamma_{F_e \to F_g}$  denotes the spontaneous decay rate on the 
$F_e \to F_g$ transition and $\alpha_{F_e \to F_g; F_{g^{'}}}$ the 
ratio between  the spontaneous decays on the $F_e \rightarrow F_{g^{'}}$ and $ F_e
\rightarrow F_g $ transitions. This ratio is given by\cite{sob}
\begin{equation}
\alpha_{F_e \to F_g; F_{g^{'}}} ={ \Gamma_{F_e \rightarrow F_{g^{'}}} \over
\Gamma_{F_e \rightarrow F_g} } =
{2 F_{g^{'}} + 1\over 2 F_g + 1}{
\left\{ \matrix{
{1\over 2} & F_{g^{'}} & {3\over 2} \cr
F_e & {1\over 2} & 1 \cr } \right\}^2   \over
\left\{ \matrix{
{1\over 2} & F_{g} & {3\over 2} \cr
F_e & {1\over 2} & 1 \cr } \right\}^2}.
 \label{alpha}
\end{equation}
The term $\alpha_{F_e \to F_g; F_{g^{'}}}$ describes the loss due to
spontaneous decay to the ground level $F_{g^{'}}$ external to  the
transition pumped by the laser. Note that the term 
$\Gamma_{F_e \to F_g}(1+\alpha_{F_e \to F_g; F_{g^{'}}})$ in Eqs.
(2) is equal to $\Gamma$ so that the description through $\alpha$
evidences the role of the spontaneous decay towards external levels. For the hyperfine
components of the D$_1$ transition the values of $\alpha$ are
\begin{mathletters} 
\begin{eqnarray}
\alpha_{ F_e = 1 \to F_g =1; F_{g^{'}}=2} &=& 5 \\
\alpha_{F_g =2 \to F_e = 2; F_{g^{'}}=1} &=& 1 \\
\alpha_{F_g =2 \to F_e = 1; F_{g^{'}}=1} &=& {1\over 5}.
\end{eqnarray}
\end{mathletters} 

In the dipole approximation the atom laser interaction has matrix
elements
\begin{equation} 
V_{e_i, g_j} = - \frac{<e_i|\vec{d}\cdot \vec{\epsilon}_x|g_j>}{2}{\cal E}.
\end{equation}

The spontaneous emission repopulation terms are\cite{happer,omont} 

\begin{equation} 
\left( {d\over dt}\rho_{g_k g_{k'}} \right)_{SE} \! \! \! \! \! \! \! =
 (2 F_e + 1)
\Gamma _{F_{e}\to F_{g}} \! \! \! \! \! \! \! \! \! \! \! \! \! \! \!
\sum_{(q,q'=-F_e,+F_e),(p=-1,1)} \! \! \! \! \! \! \! \! \! \! \! \!  
\! \! \! \! \! \! \! 
(-1)^{p-k-q'} 
\left( \matrix{
F_g & 1 & F_e \cr
-k & p & q \cr } \right)  \rho_{ e_q e_{q'}}
\left( \matrix{
F_e & 1 & F_g \cr
-q' & -p & k' \cr } \right)
\end{equation}

In order to examine the influence of the external levels on the Hanle/CPT resonance
around zero magnetic field, we have solved numerically the time-dependent OBE  with
the initial condition $\rho_{g_{i},g_{j}}(t=0) = {1 \over 8}\delta_{ij}$,
$\rho_{g_{i},e_{j}}(t=0)=\rho_{e_{i},e_{j}}(t=0)=0$. We have calculated the
time-dependent fluorescence intensity emitted from the atomic system
\begin{equation}
I(B,t) = \Gamma \sum_{m_F=-F_e, F_e} \rho_{ee}
\end{equation}
In our experiment on an atomic beam, we detect a signal proportional to the integrated
fluorescence intensity emitted by an atom interacting with the laser light during a time
$t_{f}$
\begin{equation}
I_{\rm int}(B) = \int_{0}^{t_f} I(B,t)
\end{equation}

In the case of a closed atomic system, if the time-integrated detected signal
corresponds to a long interaction times $t_{f}$, the transient initial
regime produces a negligible contribution to the overall intensity. The
contrast, between the maximum and the minimum of the emitted
fluorescence intensity, defined as in ref.\cite{ari1,ari3}, approaches hundred
percent when all the atoms are pumped into the noncoupled state. In the case of
an  open system the excited state occupation at the steady state is equal to
zero: all population is lost because of the presence of the external state. In
this case the transient regime produces the most important contribution to the
integrated emitted intensity, which exhibits a Hanle/CPT  resonance with
contrast  strongly depending on the atomic transition. Fig. 2  shows results for
$I$ versus $B$ at different interaction times and $I_{\rm int}$ versus $B$ at
$t_f=45/ \Gamma_{F_e=1\rightarrow F_g=1}$ for the $F_g=1 \to F_e=1$  hyperfine
transition of the D$_1$ line, at different values of $\alpha_{F_e \to
F_g;F_{g^{'}}}$. The case $\alpha=0$ of (a) and (b) corresponds to an ideal
close transition, the case $\alpha_{F_e=1 \to F_g=1;F_{g^{'}}=2}=5$ of (c) and
(d) corresponds to the real $F_g=1 \to F_e=1$  transition of the D$_1$ line. In
(a) and (b) for the case of a closed atomic system, where
$\alpha_{F_e \to F_g;F_{g^{'}}}=0$, the time-dependent fluorescence intensity 
exhibits a sharp and well pronounced Hanle/CPT resonance around zero magnetic
field. The contrast of $I_{\rm int}$ increases with the interaction time and results around
100 \% at larger interaction times. In (c) and (d)  a large loss
towards external states produces a reduced intensity $I$, and a smaller contrast
in the integrated intensity $I_{\rm int}$. The limiting value for the contrast observed
on $I$ is 100 percent independently of the loss rate and is reached at interaction times shorter than in the case of a closed system,
because only those atoms already in the noncoupled state contribute to the contrast, the
remaining ones been pumped into the external states. The integrated fluorescence $I_{\rm
int}$ is obtained summing up the contributions at different times, and those at earlier
times have a larger weight on the sum. Also for an open system the contrast increases with
the interaction time, reaching a value smaller than 100 \% at larger interaction times.

Fig. \ref{fig3}  shows the  results for the integrated
fluorescence intensity $I_{\rm int}$ in the case of the real open transitions of sodium,
for a choice of experimental parameters corresponding to the conditions of the
experimental investigation. The different contrast of the Hanle/CPT resonance for the
different hyperfine transitions is quite evident. A large contrast is obtained for the
$F_g=2 \to F_e=1$ transition with the smallest value of losses towards external states.
 
 The results of  Figs. \ref{fig2},\ref{fig3} show that the loss towards the external
states modifies strongly also the linewidth. In Fig. \ref{fig2}  the resonance linewidth for
the case of $\alpha_{F_e=1 \to F_g=1;F_{g^{'}}=2}=5$ is much narrower than the linewidth
for the case $\alpha = 0$. In Fig \ref{fig3} the real hyperfine transitions are directly
compared and once again open transitions with a larger loss towards external states have
a more narrow resonance. The comparison between different hyperfine transitions is not
straightforward because the matrix element $V_{e,g}$ of the atom-laser
interaction between excited and ground states depends on the hyperfine levels, and
furthermore in the system $F_g=2\to F_e=1$ two noncoupled states are present. However 
the CPT analysis for a closed symmetrical $\Lambda$ system performed in the basis of the
coupled and noncoupled states is very usedul for the interpretation. In
that analysis \cite{ari1,ari3}, in absence of ground state relaxation, at the steady
state the linewidth of the CPT resonance is determined by the loss rate $\Gamma^\prime$ of
the coupled state, with  $\Gamma^{\prime} = V_{e,g}/(\hbar^2\Gamma_{\rm exc})$,
 $\Gamma_{\rm exc}$ being the excited state lifetime. In Fig. \ref{fig2}, the main
difference between the closed, in (a) and (b), and open, in (c) and (d), systems is the
excited state lifetime, with  $\Gamma_{\rm exc}=\Gamma_{1 \to 1}$ in the first
case, and $\Gamma_{\rm exc}=\Gamma_{1 \to 1}\alpha_{1 \to 1;2}$
in the second case. An increase in that lifetime produces a smaller $\Gamma^\prime$
and a more narrow CPT resosnce, as observed in the figure for both $I$ and $I_{\rm int}$.
It should be noted that in an open system the coupled-noncoupled approach applied to
states without population because of the optical pumping towards external states is not
really meaningful. 

The narrow linewidths obtained in Fig. \ref{fig3} have a different explanation, because
for the real D$_1$ line hyperfine transitions  the excited states have the same lifetime
$\Gamma$, whichever Zeeman/hyperfine component. The narrow resonances are obtained
in the curve of $I_{\rm int}$ versus $B$ and the contribution of $I$ at different
interaction times should be considered. They  arise because of the simultaneous action of
the pumping into the noncoupled nonabsorbing state and of the optical pumping towards
external states. The optical pumping towards external states is less efficient whenever the
optical pumping in the noncoupled states is more efficient: in the noncoupled state the
atomic wavefunction hsa no contribution of excited state, whence the atom
does not decay towards the external states. Thus the narrow CPT
resonance, being produced by the contribution of only those atoms remaining in the
noncoupled state and nondecaying towards the external states, is a consequence of an atomic
selection.

\section{experimental set-up}

The experimental setup is shown in Fig. \ref{fig4}. We used a single mode CW dye laser;
the light polarization was linear and the propagation direction orthogonal to
the thermal sodium atomic beam. An external magnetic field was applied
parallel to the laser beam propagation direction. Magnetic field Helmholtz
coils, applied around the atom-laser interaction region,  were driven by a
programmable power supply in order to produce a linear scan  of  the magnetic
field.  The earth magnetic field was compensated to better than $0.05$ G by 
additional pairs of Helmholtz coils.
By tuning the laser to different hyperfine optical transitions of the sodium
D$_1$ line,  we excited  different atomic configurations.  The laser frequency
was tuned to the center of the homogeneous hyperfine absorption profile.
The fluorescence signal emitted by the sodium atoms at right angles with
respect to the directions of the laser field propagation and of the magnetic field
was detected through a photomultiplier and recorded by means of a lock-in
amplifier and a standard data acquisition system. In our experiment the
interaction time was of the order of  few $\mu $s, to be compared to the
$3P_{1/2}$ excited state lifetime  of sixteen $n$s.

\section{experimental results and comparison with theory}

Examples of the  time integrated fluorescence signals recorded on the different
hyperfine transitions are shown in Fig. \ref{fig5}. Three of the four
transitions on the D$_1$ line exhibit the Hanle/CPT typical dip. The transition
$F_g=1 \rightarrow F_e=2$ does not  exhibit any dip because there are no
noncoupled states into which the atom could  be pumped. The dip contrasts for
the other three transitions, $F_g=1 \rightarrow F_e=1, F_g=2 \rightarrow F_e=1,
F_g=2 \rightarrow F_e=2$, are very  different. The dip for the transition $F_g=2
\rightarrow F_e=1$ is very pronounced, and the corresponding contrast is large.
The contrasts of the dips associated with the 
transitions $F_g=1 \rightarrow F_e=1$  and $F_g=2 \rightarrow F_e=2$
are smaller. The observed behaviours correspond to those predicted by the analysis and
shown in Fig. \ref{fig2}. The hyperfine transitions with a larger loss factor $\alpha$
towards external states present a more narrow Hanle/CPT resonance. However it should be
noted  that different hyperfine transitions cannot be directly compared
as already discussed.

The FWHM linewidth and the contrast of the Hanle/CPT resonance for the various
hyperfine transitions have been studied as a function of the laser intensity.  The
experimental results are shown in Fig. \ref{fig6} and \ref{fig7}a.  The experimental results
of Figs. \ref{fig6} and \ref{fig7}a for the linewidth and contrast of the Hanle/CPT resonance have been
 compared to the theoretical numerical analysis. The theory predicts an
increase of the linewidth and contrast with the applied laser intensity, as observed in the
experiment. However the maximum contrast achieved in our experimental observations is lower
than that predicted by the theory. For instance for the $F_g=2 \to F_e=1$
hyperfine transition, the maximum theoretical contrast is sixtyseven percent, while the
measured value of the maximum contrast on that transition is around fifty percent. In
order to examine more carefully the comparison between theory and experiments, our
theoretical analysis has been applied also to the Hanle/CPT measurements published by
Picqu\'e \cite{picque} in an experimental configuration very similar to the present one, on
the $F_g=2 \to F_e=1$ hyperfine transition of the sodium  D$_1$ line as a function of the
applied laser field intensity. The contrast of the Hanle/CPT resonances, as derived from the
five curves published in \cite{picque}, and the results of a theoretical analysis for the
interaction times of that reference are reported in Fig. \ref{fig7}b. In this case the
agreement between our theory and the experimental results is very good,
confirming the validity of the theoretical approach.  The experiment of
\cite{picque} reported a compensation of the stray magnetic fields within 10 mG
over the whole atom-laser interaction volume. Thus  our failure to reproduce the
theoretical contrast is produced by an imperfect compensation of the magnetic
field in our apparatus. More precisely  owing to the presence of some
magnetic field gradient present in the atom-laser interaction region, the
magnetic field could not be compensated over the whole interaction region through
our Helmholtz coils.  

For the D$_1$ excitation on sodium atoms none of the hyperfine optical transitions is
closed. However experimentally a closed-like situation can be realized  through the
application of a repumping laser which compensates the losses term $\dot{\rho}_{e_i
e_i}= - \alpha_{F_e \to F_g;F_{g^{'}}}\Gamma_{F_e \to F_g} \rho_{e_ie_i}$ for the
population decay. We examined, for a given transition, the dependence of 
contrast and linewidth  on the population loss rate, by  applying a  repumping
laser to partially compensate the losses towards external states.   Varying  the
intensity of the repumping laser it was possible to study the features  of the
CPT resonance as a function of the rate of population losses. Obviously the
repumping transition should be chosen in order to not produce an additional
noncoupled coherent superposition. In the case of the sodium D$_1$ line the
most favourable transition for this study is the  $F_g=2 \rightarrow F_e=2$ 
transition because the transition $F_g =1 \rightarrow F_e =2$,  without
noncoupled state,  as verified by the record in Fig. \ref{fig5}d, can be used
for repumping. The CPT results in presence of a repumping laser are reported in
Fig. \ref{fig8}: it is clearly visible that the contrast and the width of the CPT
resonance  increase for increasing power of the repumping laser, i.e., for a
decreasing rate of population losses. These experimental results can be compared to the
theoretical ones of Fig. \ref{fig2}. The presence of a repumping laser, compensating for
the losses towards external states, is equivalent to a longer lifetime of the excited
state. Thus starting from the configuration of an open system as in Fig. \ref{fig2}(b) in
absence of the repumping laser, its application produces an effective closed system as in
Fig. \ref{fig2}(a). It should be noted that the repumping laser does not compensate for the
decay rate of the optical coherences. However after preparation of the atoms in the
noncoupled states, the optical coherences are zero and their decay is not relevant for the
atomic preparation.

\section{conclusions}

The CPT phenomenonon the sodium $D_{1}$
line has been investigated by means of an Hanle effect configuration comparing  
different level schemes ($\Lambda$ and $M$) and studying the 
influence of the losses  towards external states. Different
atom-laser interaction schemes involving as ground states the sublevels of the same
hyperfine component have been considered. On the sodium D$_1$ line the
relevant transition  are   $F_g=1\rightarrow F_e=1$,
$F_g=2\rightarrow F_e=1$, $F_g=2\rightarrow F_e=2$. 
As original contribution of the present investigation we have investigated CPT in
atomic configurations not closed for spontaneous emissions decay. We have
demonstrated that CPT may be realized on those open transition with an efficiency
decreasing with the amount of spontaneous emission towards external states. Because
in the D$_1$ line the different level configurations correspond  to different rates of 
population losses, the features, i.e., width and contrast, of the Hanle/CPT resonance
depend on the explored transition. A careful study of the dependence of the width and
contrast of the  Hanle/CPT dip on the interaction parameters has been performed.  A
numerical analysis of the density matrix equations confirms most experimental
observations, except that the maximum contrast is lower than the expected one because
of the not perfect compensation of the magnetic field in the experiment. However our
numerical analysis is in agreement with the contrast measured in a previous experiment
\cite{picque} where a very good magnetic field compensation was applied.  The
experimental data show that the contrast of the Hanle/CPT dip is less than 100 \%. It
depends strongly on the population loss rate and it can be controlled by means of a
repumping laser  that compensates the rate of population losses towards the external
levels. The contrast of the Hanle/CPT resonance
 is only weakly affected by the small Doppler broadening associated to the
atomic beam. 

Note that in the regime of very large saturation, the
power broadening  of the spectral lines does not allow the interpretation 
of the experiment as a pure $F_g\rightarrow F_e$ transition with losses
on $F_{g^{'}}$. Let us consider for example  the case of $F_g=1 \rightarrow 
F_e=1$ ($\Lambda$-scheme). At large laser saturation the absorption  lines 
$F_g=1\rightarrow F_e=1$ and  $F_g=1\rightarrow F_e=2$ are 
excited simultaneously. Thus we don't have a pure 
$\Lambda$-system but a $\Lambda+W$ system where the  states are not 
completely  dark for the laser absorption.

\section{Acknowledgments}

This work has been partially supported by the Austrian Science
Foundation No. S 6508.

\newpage

\begin{figure}
\caption{Different CPT schemes that can  be established on the hyperfine components of 
the sodium D$_1$ line through excitation by  $\sigma_{+},\sigma_{-}$ laser lights.} 
\label{fig1} 
\end{figure}

\begin{figure}
\caption{Theoretical results for the time dependent intensity $I$ versus time $t$ and
integrated fluorescence intensity $I_{\rm int}$ at $t_f =45$ for the $F_g=1\rightarrow
F_e=1$ hyperfine transition in correspondence of different values of the loss parameter
$\alpha_{F_e=1\rightarrow F_g=1,F_{g^{'}}=2}$. In (a) and (b)  $\alpha_{F_e=1\rightarrow
F_g=1,F_{g^{'}}=2}=0$, i.e. a hypothetical closed transition; in (c) and (d)
$\alpha_{F_e=1\rightarrow F_g=1,F_{g^{'}}=2}=5$ as for the real hyperfine transition of the
D$_1$ line. The time $t$ and $t_f$ are measured in units of $1/ \Gamma_{F_e=1\rightarrow
F_g=1}$. The laser intensity is 26 mW/cm$^2$.}     
\label{fig2}   
\end{figure}        

\begin{figure}
\caption{Theoretical results for the integrated fluorescence intensity $I_{\rm
int}$ at $t_f =45/ \Gamma_{F_e=1\rightarrow
F_g=1}$ for the real (open) transitions of the sodium D$_1$ line,
calculated for the laser intensity of 26 mW/cm$^2$. (a) corresponds to the
transition $F_g=1\rightarrow F_e=1$, (b) to $F_g=2\rightarrow F_e=1$,  and (c)
to $F_g=2\rightarrow F_e=2$.}  
\label{fig3}  
\end{figure}

\begin{figure}
\caption{Experimental setup for the investigation of the Hanle/CPT  resonance on a sodium
atomic beam for the different hyperfine transitions on the D$_1$ line.} 
\label{fig4} 
\end{figure}

\begin{figure}
\caption{Experimental results for the fluorescence intensity $I_{\rm int}$ as a function of
the magnetic field $B$ on the different hyperfine transitions with no repumping laser. The
interaction time is $\simeq 4.5 \mu$s. Laser intensity $I_{L} =$ 26 mW/cm$^2$ in all the data sets.} 
\label{fig5} 
\end{figure}

\begin{figure}
\caption{Experimental results for the FWHM linewidth of the Hanle/CPT resonance for the
different hyperfine transitions as a function of the laser intensity. Black triangles
 correspond to laser excitation on the transition $F_g=2\rightarrow F_e=2$, while
open squares  to  $F_g=1\rightarrow F_e=1$ and closed circles to $F_g=2\rightarrow
F_e=1$. Typical error bars are marked.} 
\label{fig6} 
\end{figure}

\begin{figure}
\caption{In (a) experimental results for the contrast of the Hanle/CPT
resonance for the different hyperfine transitions  as a function of the 
laser intensity. Black triangles correspond to the transition 
$F_g=2\rightarrow F_e=2$, open squares to  $F_g=1\rightarrow
F_e=1$ and closed circles to $F_g=2\rightarrow F_e=1$. 
Typical error bars are marked.
In (b) experimental results for the $F_g=2\rightarrow F_e=1$ transition 
from ref. \protect\cite{picque} for the contrast of the Hanle/CPT resonance  
as a function of the laser intensity. Laser intensity derived from
\protect\cite{picque}; error bar of the contrast extimated  on the figures of
that reference. }  \label{fig7}  \end{figure} 

\begin{figure}
\caption{Experimental results for the fluorescence intensity as a function of the applied
magnetic field, with one laser tuned to the transition $F_g=2\rightarrow F_e=2$ to establish
the M-scheme;  another laser tuned to the transition $F_g=1\rightarrow F_e=2$ to repump the
atoms. The intensity of the laser used to establish the M-scheme was $I_L= 250$ mW /
cm$^2$. The intensity of the repumping laser was in (a) $I_R= 0$, in (b) $I_R =$ 20
mW/cm$^2$, (c) $I_R= $200 mW/cm$^2$.}   
\label{fig8}  
\end{figure}        

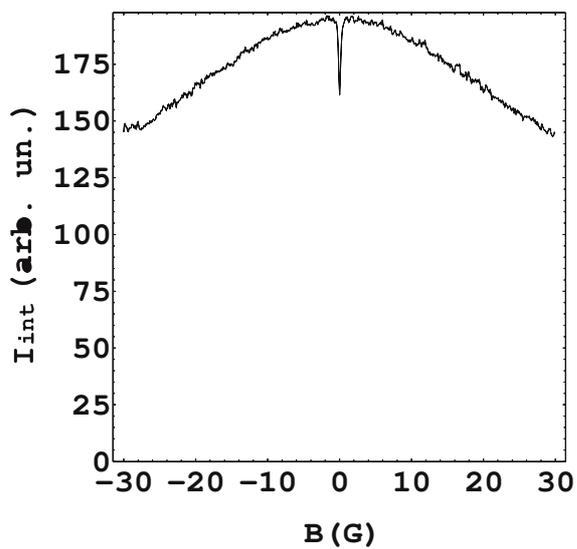
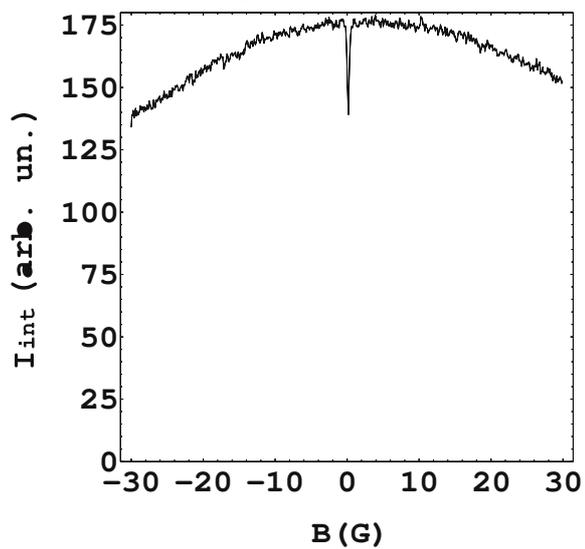
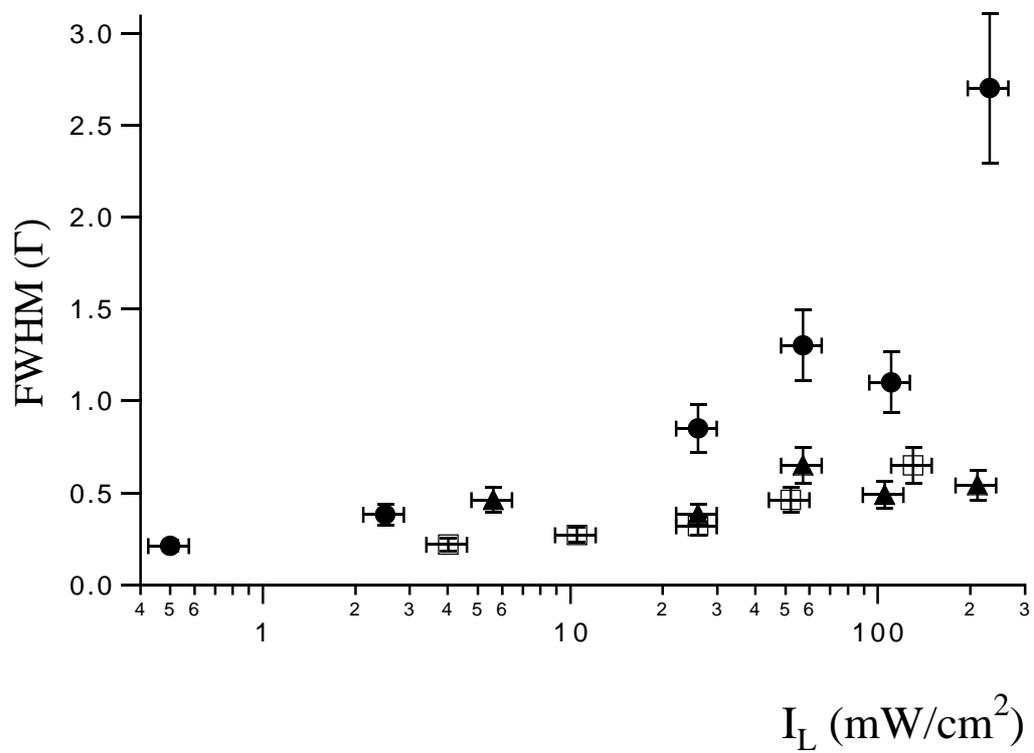
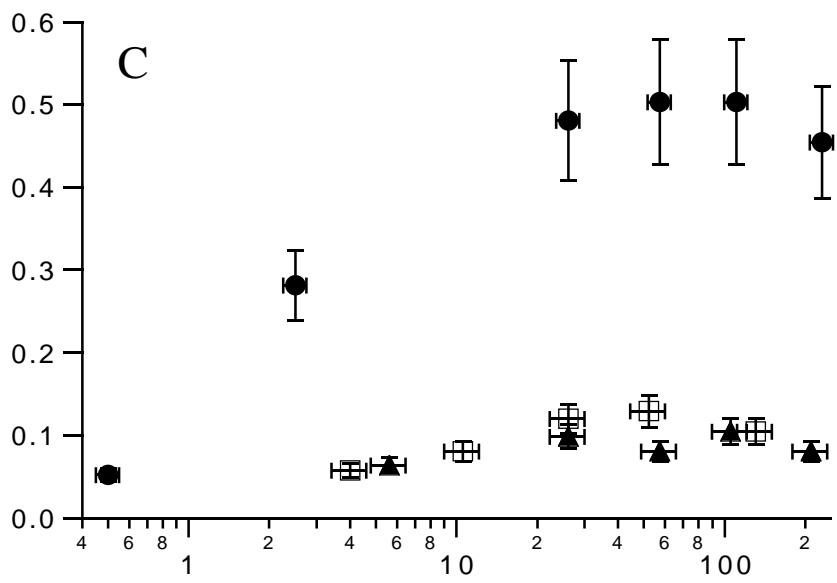
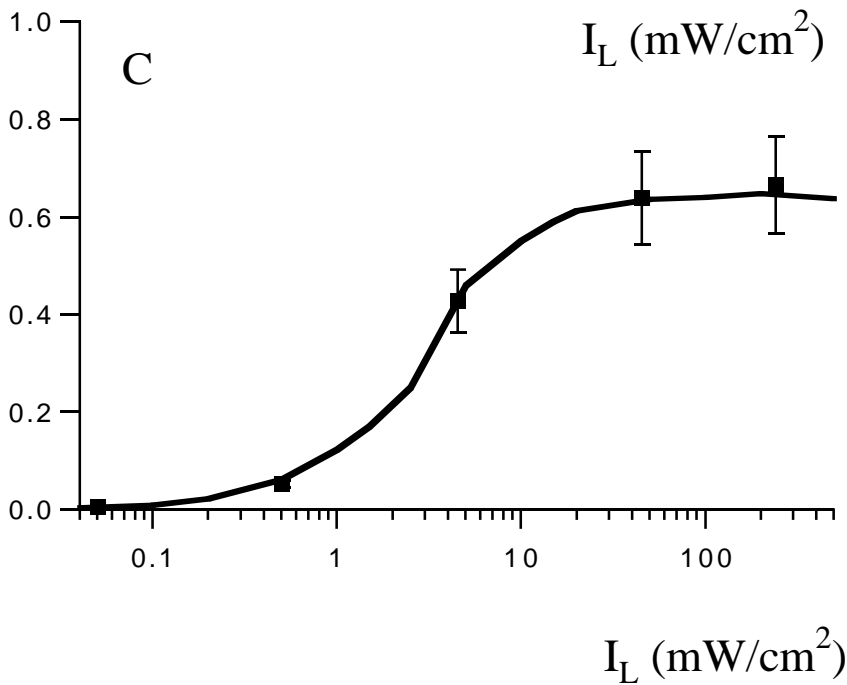
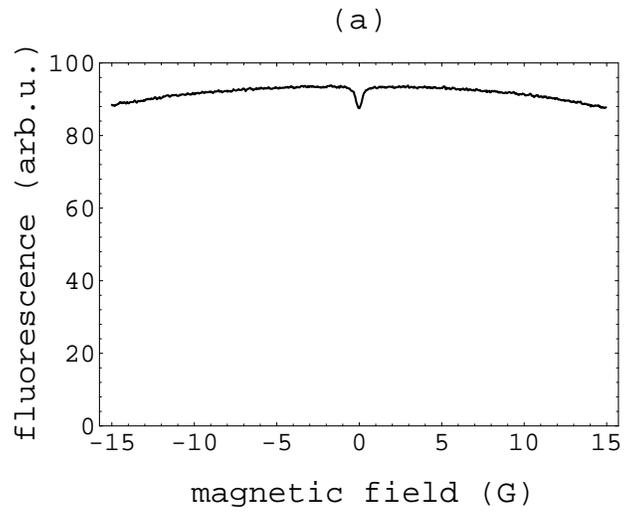
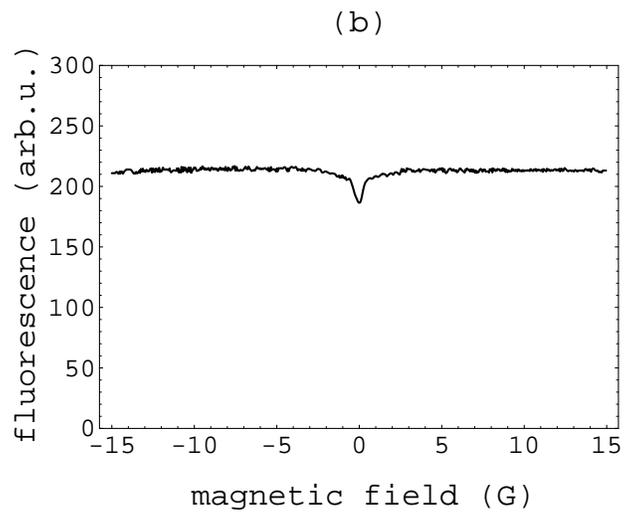
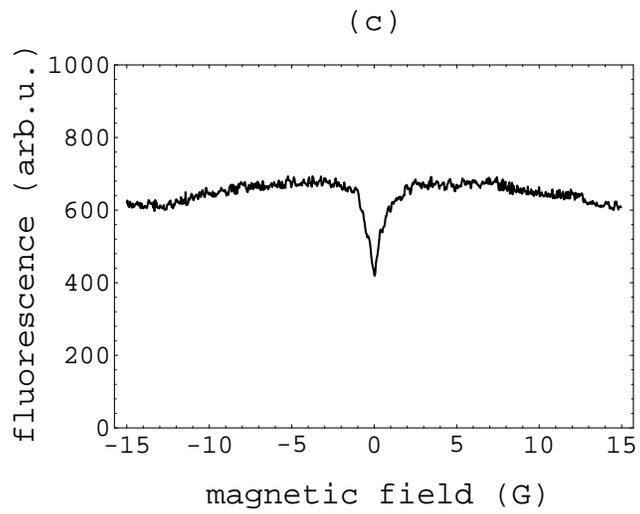
\end{document}